\documentclass[12pt,a4paper,final]{iopart}

\newcommand{\ket}[1]{\left| #1 \right \rangle}

\newcommand{\sspec}{^{1}\Sigma^{+}}

\usepackage{color}
\usepackage{iopams}
\usepackage{graphicx}
\usepackage[breaklinks=true,colorlinks=true,linkcolor=blue,urlcolor=blue,citecolor=blue]{hyperref}

\begin{document}

\title[]{Formation of molecular ions by radiative association of cold trapped atoms and ions}

\author{Humberto da Silva Jr.$^{1}$}
\address{$^1$Laboratoire Aim$\acute{e}$ Cotton, CNRS/Universit$\acute{e}$ Paris-Sud/ENS Cachan, Orsay Cedex, France}

\author{Maurice Raoult$^{1}$}
\address{$^1$Laboratoire Aim$\acute{e}$ Cotton, CNRS/Universit$\acute{e}$ Paris-Sud/ENS Cachan, Orsay Cedex, France}

\author{Mireille Aymar$^{1}$}
\address{$^1$Laboratoire Aim$\acute{e}$ Cotton, CNRS/Universit$\acute{e}$ Paris-Sud/ENS Cachan, Orsay Cedex, France}

\author[cor1]{Olivier Dulieu$^{1}$}
\address{$^1$Laboratoire Aim$\acute{e}$ Cotton, CNRS/Universit$\acute{e}$ Paris-Sud/ENS Cachan, Orsay Cedex, France}
\eads{\mailto{olivier.dulieu@u-psud.fr}}

\begin{abstract}
Radiative emission during cold collisions between trapped laser-cooled Rb atoms and alkaline-earth ions (Ca$^+$, Sr$^+$, Ba$^+$) and Yb$^+$, and between Li and Yb$^+$, are studied theoretically, using accurate effective-core-potential based quantum chemistry calculations of potential energy curves and transition dipole moments of the related molecular ions. Radiative association of molecular ions is predicted to occur for all systems with a cross section two to ten times larger than the radiative charge transfer one. Partial and total rate constants are also calculated and compared to available experiments. Narrow shape resonances are expected, which could be detectable at low temperature with an experimental resolution at the limit of the present standards. Vibrational distributions are also calculated, showing that the final molecular ions are not created in their ground state level.
\end{abstract}

\vspace{2pc}

\section{Introduction}

One of the novel developments of ultracold matter research is exemplified by the experiments aiming at merging a cold atom trap and a trap of laser-cooled ions. Such hybrid setups could offer the opportunity to study collisional dynamics in the quantum $s$-wave scattering regime associated to the long-range interaction between the ion and the atom varying as $R^{-4}$ (where $R$ is the ion-atom distance) \cite{cote2000a,idziaszek2009,gao2011,krych2011}, just like it has been extensively studied for ultracold neutral atom-atom collisions (see for instance Ref.~\cite{quemener2012}). These hybrid devices have also been proposed to emulate solid-state physics where a band structure in the fermionic atoms -thus simulating electrons- could be induced by periodic structure generated by the laser-cooled ions \cite{bissbort2013}. Other prospects concern the implementation of a Josephson junction through a pair of cold atoms controlled by a single ion \cite{gerritsma2012,joger2014}, the immersion of a single ion into a Tonk-Girardeau gas to generate an ionic density "`bubble"' \cite{goold2010}, or the elaboration of an atom-ion quantum gate \cite{doerk-bending2010}. The variety of combinations of laser-cooled alkali-metal atoms and alkaline-earth (or rare earth) atomic ions allows to explore the limit of sympathetic cooling between them depending on their mass ratio \cite{krych2011,cetina2012,nguyen2012,tomza2015}.

Also, a wealth of cold molecular ions species could be created, opening the way to a rich chemistry at temperatures of a few millikelvin, or less \cite{willitsch2008}. Following the theoretical prospect of Ref.~\cite{makarov2003} with cold Na trapped in a magnetooptical trap (MOT) and laser-cooled Ca$^+$ ions in a Paul trap (then followed by further developments \cite{smith2014}), other groups have carried out pioneering experiments with various combinations of atoms and atomic ions: Yb atoms with Yb$^+$ ions \cite{grier2009}, Rb atoms with Ca$^+$ \cite{hall2011,hall2013a} and Ba$^+$ \cite{hall2013b} ions, and Ca atoms with Yb$^+$ ions \cite{rellegert2011} or Ba$^+$ ions \cite{sullivan2012}. Rubidium atoms can also be confined in a magnetic trap in (or close to) the Bose-Einstein condensation (BEC) regime thus ensuring with a density larger than in a MOT, and then interacting with a few Yb$^+$ ions \cite{zipkes2010,zipkes2010a,zipkes2011,ratschbacher2012}. Similarly, an optical dipole trap of Rb atoms has been merged in a Paul trap containing a few Ba$^+$ atoms \cite{schmid2010,schmid2012} (or with the combination Li/Ca$^+$ \cite{haze2013}). When atomic ions like Rb$^+$ or Na$^+$ cannot be laser-cooled, they can be sympathetically cooled by another species in the Paul trap in the presence of trapped Rb atoms \cite{haerter2012,schmid2012} or Na atoms \cite{sivarajah2012}, or they can be created \textit{in situ} inside the Paul trap by photoionization of trapped Rb atoms \cite{ravi2011,ravi2012,lee2013}. To be complete, sympathetically-cooled molecular ions created from an external source have been utilized in two remarkable experiments: one has demonstrated the vibrational quenching of BaCl$^+$ ions by laser-cooled Ca atoms \cite{rellegert2013}, and another one has recorded a particularly large charge-exchange rate between N$_2^+$ ions and laser-cooled Rb atoms \cite{hall2012}. A comprehensive review of most of these experiments can be found in Ref.~\cite{haerter2014}.

Elastic and inelastic collisions as well as elementary reactions like charge transfer between the trapped ions and the trapped atoms have been studied in these experiments. Due to the residual micromotion of the trapped ions, these studies have been limited to energies larger than a few millikelvin, well beyond the $s$-wave quantum regime. It has been recognized in several of these experiments (see for instance Ref.~\cite{hall2011}) that the cooling and/or trapping light most often assist the reaction through the excitation of one of the colliding partners, inducing large rates described by the Langevin regime close to the unitarity limit.

Among all these experiments, it is striking that so far, the direct observation of molecular ions resulting from the association of a cold atom and a cold ion has been reported only in two cases, namely RbCa$^+$ \cite{hall2011,hall2013a} and RbBa$^+$ \cite{hall2013b}. These observations have been assigned to the formation of ions by radiative association (RA), \textit{i.e.} Rb + (Alke)$^+ \,\rightarrow$ Rb(Alke)$^+ + h\nu$ (where (Alke) = Ca, Ba), and $h\nu$ is the energy of the photon emitted during the collision in order to stabilize the molecular ion \cite{hall2011,hall2013a,hall2013b}. The RA process competes with radiative charge transfer (RCT) Rb + (Alke)$^+ \,\rightarrow$ Rb$^+ \,+$ (Alke) $+h \nu$ when (Alke) = Ba, and is dominated by non-radiative charge transfer (NRCT) in the Rb/Ca$^+$ case \cite{tacconi2011}. The RA of CaYb$^+$ has been invoked -- but molecular ions have not been directly detected -- in Ref.~\cite{rellegert2011} to explain the measured fast reaction rate, but not confirmed by a subsequent theoretical investigation \cite{zygelman2014}.

In order to clarify the possible molecular ion formation processes and their efficiency in such merged ion and atom traps, we achieved in this paper a systematic and consistent analysis of the inelastic collisions of cold Rb/(Alke)$^+$ pairs in their ground state considering the two competitive channels of RA and RCT. Here the notation (Alke) refers to the alkaline-earth atoms Ca, Sr, Ba. For convenience the lanthanide atom Yb, with its closed $4f$ shell and its external $6s^2$ shell, will be included in this (Alke) notation. The Li-Yb$^+$ complex is also studied as its significantly different mass ratio motivated the analogy for using this combination to emulate solid-state physics \cite{bissbort2013}. Occasionnally, the Rb and Li alkali-metal atoms will be referred to as the (Alk) species. After recalling the basics of the derivation of expressions for the cross section and the rates (Section \ref{sec:theory}), we describe the electronic structure (potential energy curves, transition and permanent electric dipole moments) of the relevant molecular ions (Alk-Alke)$^+$ using accurate quantum chemistry calculations based on the representation of the colliding partners with effective core potentials (ECP) including core polarization potentials (Section \ref{sec:structure}) which are compared to other determinations when available. We present our results for both RA and RCT cross sections and rates, emphasizing then on the presence of narrow shape resonances induced by the centrifugal barrier in the entrance channel,  and on the vibrational distribution of the molecular ions, which are not expected to be created in their ground level (Section \ref{sec:rates}). We finally discuss some prospects about possible explanation for the absence of molecular ions in several of the experiments quoted above (Section \ref{sec:discussion}).

In the rest of the paper atomic units of energy (1~a.u. $=2\times R_{\infty}$ where $R_{\infty}$ is the Rydberg constant), distance (1~a.u. = $a_0$ where $a_0$ is the Bohr radii), and dipole moment (1~a.u. = 2.541 580 59~D) will be used, except otherwise stated. The collision energies will usually be expressed in units of (milli-)Kelvin, $E_{coll} \equiv E_{coll}/k_{b}$, where $k_{b}$ is the Boltzmann constant.


\section{Theoretical approach for RA and RCT}
\label{sec:theory}

The spontaneous emission of a photon during a collisional process is modeled by a well-established formalism. It is a special case of a well-known semiclassical matter-radiation interaction scheme of absorption and emission. A $1^{st}$-order time-dependent perturbation approach expresses the atomic or molecular transition between an initial state $\ket{i}$, with energy $E_{i}$ in the instant $t_{i}$ and a final state $\ket{f}$, with energy $E_{f}$ at $t > t_{i}$, induced by the perturbation of the electromagnetic field $H' = H'(t > t_{i})$. Such a methodology has been successfully applied to the description of the radiative association \cite{zygelman1990,gianturco1996,gianturco1997,zygelman1998,zygelman2014} and of the radiative charge transfer in atom-ion collisions, for elementary systems such as Li(2s) + H$^{+}$ \cite{stancil1996}. In the following we only recall the main steps, and a full description can be found in the given references and the references therein.

In the collisions between ground-state (Alke)$^+$($ns$) ions ($n=4,5,6,6$ for Ca, Sr, Ba, Yb, respectively) and Rb($5s$), or between Yb$^+$($6s$) and Li($2s$) atoms, the  entrance channel is associated with a single adiabatic Born-Oppenheimer (ABO) potential energy curve (PEC) $V_i(R)$ of $^1\Sigma^+$ symmetry which actually corresponds to the first excited electronic state A$^1\Sigma^+$ of the (Alk-Alke)$^+$ molecular ion (Fig.~\ref{fig:pec}a). Note that this may not be the case for this class of mixed species, as for instance in LiCa$^+$ and LiBa$^+$ where the (Alke)$^+$($ns$)+Li($2s$) dissociation limit is the lowest one. The spontaneous emission during the collision thus leads either to the dissociation continuum (for RCT) or to the bound rovibrational level manifold (for RA) of the PEC $V_f(R)$ of the (Alk-Alke)$^+$ ground electronic state X$^1\Sigma^+$ correlated to the lowest  (Alke)($ns^2\,^1S$) + (Alk)$^+$ asymptote. Note that below the entrance channel there is an additional asymptotic limit only in RbCa$^+$, namely  Rb$^+$ + Ca($4s4p\,^3P$), to which the PEC lowest $^3\Pi$ state is correlated and which is coupled to the A state by spin-orbit interaction, thus inducing non-adiabatic transition, \textit{i.e.} NRCT \cite{tacconi2011}. This feature is not treated in the present paper. Also, as illustrated for instance in our previous work on alkali-metal atom and Sr$^+$ compounds \cite{aymar2011}, the lowest $^3\Sigma^+$ PEC is correlated to the entrance channel (Alke)$^+$($ns$) + (Alk)($n's$), but cannot give rise to spontaneous emission to the X state, as long as no second-order spin-orbit coupling is introduced. The hyperfine structure of the colliding partners -- which has been found to be an important feature in Ref.~\cite{zipkes2010} -- is neglected, so that the entrance channel A$^1\Sigma^+$ is assumed to be populated with a statistical weight of $p=1/4$.

Following our previous investigation \cite{ayouz2011} and similar ones (see for instance Refs. \cite{zygelman1988,gianturco1997}, the total spontaneous emission cross section during the collision at energy $E_{coll}$ is formulated as the sum of two contributions from the RCT and RA processes
\begin{eqnarray}
\sigma^{RCT}(\epsilon_{i}) &=& p \frac{8\pi^2}{3c^3}\frac{1}{k_i^2}  \sum_{J=0}^{\infty} \int_{0}^{\epsilon_{f}^{max}} [\left(\omega^3_{i,f} J | \langle J-1, \epsilon_{f}|D(R)|\epsilon_{i},J\rangle |^{2} \right) \nonumber \\
&+&\left(\omega^3_{i,f}(J+1) |\langle J+1, \epsilon_{f}|D(R)|\epsilon_{i},J\rangle |^{2}\right) ]d\epsilon_{f}
\label{eq:csrct}
\end{eqnarray}
\begin{eqnarray}
\sigma^{RA}(\epsilon_{i}) &=& p \frac{8\pi^2}{3c^3}\frac{1}{k_i^2}  \sum_{J=0}^{\infty} \sum_{v=0}^{v_{max}}[ \left( \omega^3_{iJ,v(J-1)} J | \langle J-1,v|D(R)|\epsilon_{i},J\rangle |^{2}\right) \nonumber \\
&+&\left(\omega^3_{iJ,v(J+1)}(J+1)|  \langle J+1, v|D(R)|\epsilon_{i},J\rangle |^{2}\right) ]
\label{eq:csra}
\end{eqnarray}
All the quantities in these equations are expressed in atomic units. In Eq.~(\ref{eq:csrct}) the integral is limited to the largest possible energy ${\epsilon_{f}^{max}}$ in the exit channel, while in Eq.~(\ref{eq:csra}) the summation is limited to the uppermost vibrational level $v_{max}$ of the X state. The initial ket $|\epsilon_i,J\rangle $ with energy $\epsilon_{i}=E-V_i(R=\infty)$ and associated wavenumber $k_{i}=\sqrt{2\mu \epsilon_{i}}$ ($\mu$ is the reduced mass of the (Alk-Alke)$^+$ system) represents a partial-wave component $J$ of an energy-normalized continuum wavefunction of the two nuclei. As only $^1\Sigma^+$ states are involved, $J$ is the total angular momentum in the entrance channel, and $J'=J\pm1$ are the two possible allowed total angular momentum quantum numbers in the exit channel. In both equations above, the summation on $J$ is actually limited for each collisional energy to the maximal value for which the induced rotational barrier in the entrance channel prevents the collision to occur. The transition electric dipole moment (TEDM) function $D(R)$ is represented in Fig.~\ref{fig:pec}c for each system.

For RCT the final state is represented by $|\epsilon_f,J\pm 1\rangle$ with energy $\epsilon_{f}=E-V_f(R=\infty)-\hbar \omega_{if}$, where $\omega_{if}$ is the energy of the emitted photon. Equation~(\ref{eq:csrct}) contains the transition dipole moment between $|\epsilon_i,J\rangle $ and $|\epsilon_f,J'\rangle $
\begin{equation}
\langle J', \epsilon_{f}|D(R)|\epsilon_{i},J\rangle=\int_{0}^{\infty} F_{J'}^{f}(\epsilon_{f},R) D(R) F_{J}^{i}(\epsilon_{i},R)~ dR
\label{eq:dipcc}
\end{equation}
involving two energy-normalized continuum wave functions $F_{J}^{i}(\epsilon_{i},R)$ and $F_{J'}^{f}(\epsilon_{f},R)$. At large distances they behave like
\begin{equation}
F_{\ell}(\epsilon,R)\sim\sqrt{\frac{2\mu}{\pi k}} \sin(kR-\ell \frac{\pi}{2}+\delta_{\ell})
\end{equation}
For RA the final state $|v,J\rangle $ is a bound rovibrational level of the X$^1\Sigma^+$ ground electronic state of the (Alk-Alke)$^+$ molecular ion. The corresponding transition dipole moment is
\begin{equation}
\langle J', v|D(R)|\epsilon_{i},J\rangle=\int_{0}^{\infty} \chi_{vJ'}^{f}(R) D(R) F_{J}^{i}(\epsilon_{i},R)~ dR
\label{eq:dipcb}
\end{equation}
where $\chi_{v}$ is a radial rovibrational wave function of the molecule normalized to unity.

Our calculations cover an energy range in the entrance channel from 1~mK up to 80~mK, currently accessible in most experiments. This requires to consider $J$ values in the range [15;40], [20;50], [20;50], [20;55], [5;15]  for RbCa$^+$, RbSr$^+$, RbCa$^+$, RbYb$^+$, and LiYb$^+$ respectively, in order to obtain converged cross sections. The integrations over the internuclear distance $R$ are performed with the Milne phase-amplitude method \cite{milne1930,korsch1977}. Such high $J$ values and low initial temperature impose to propagate out the continuum wave function in the entrance channel up to large $R$ values, namely 20000~a.u. This ensures that the  outward propagation is stopped well within the classically allowed region situated at long range beyond the centrifugal barrier. The  upper limit $\epsilon_{f}^{max}$ = 1000~cm$^{-1}$ of the relative energy in the integral of Eq.~(\ref{eq:csrct}) and the integration step  $d\epsilon_{f}$ = 1cm$^{-1}$ are taken to ensure the integral convergence.

\begin{figure}[!htbp]
\centering
\includegraphics[scale=0.28, angle=0]{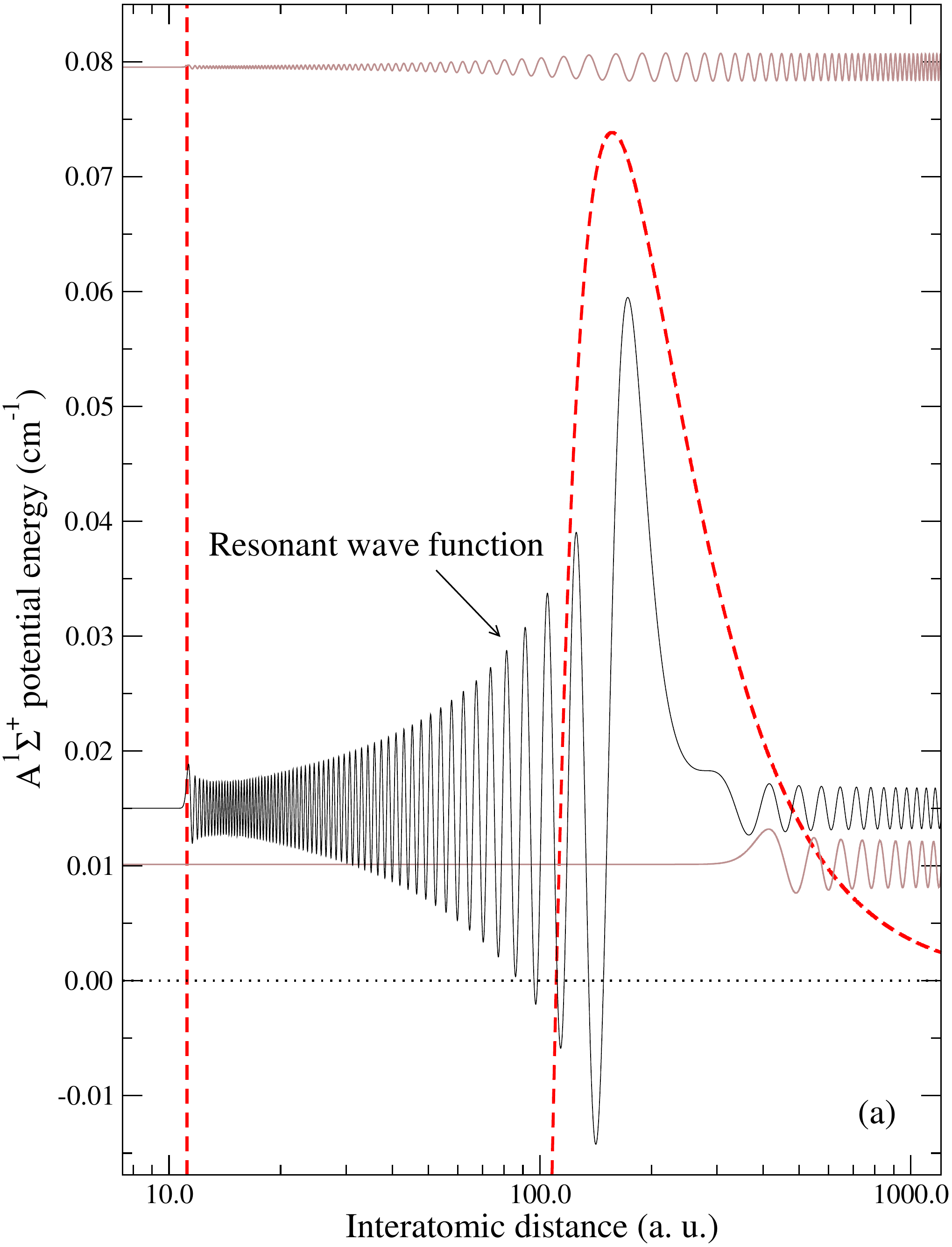}
\includegraphics[scale=0.74, angle=0]{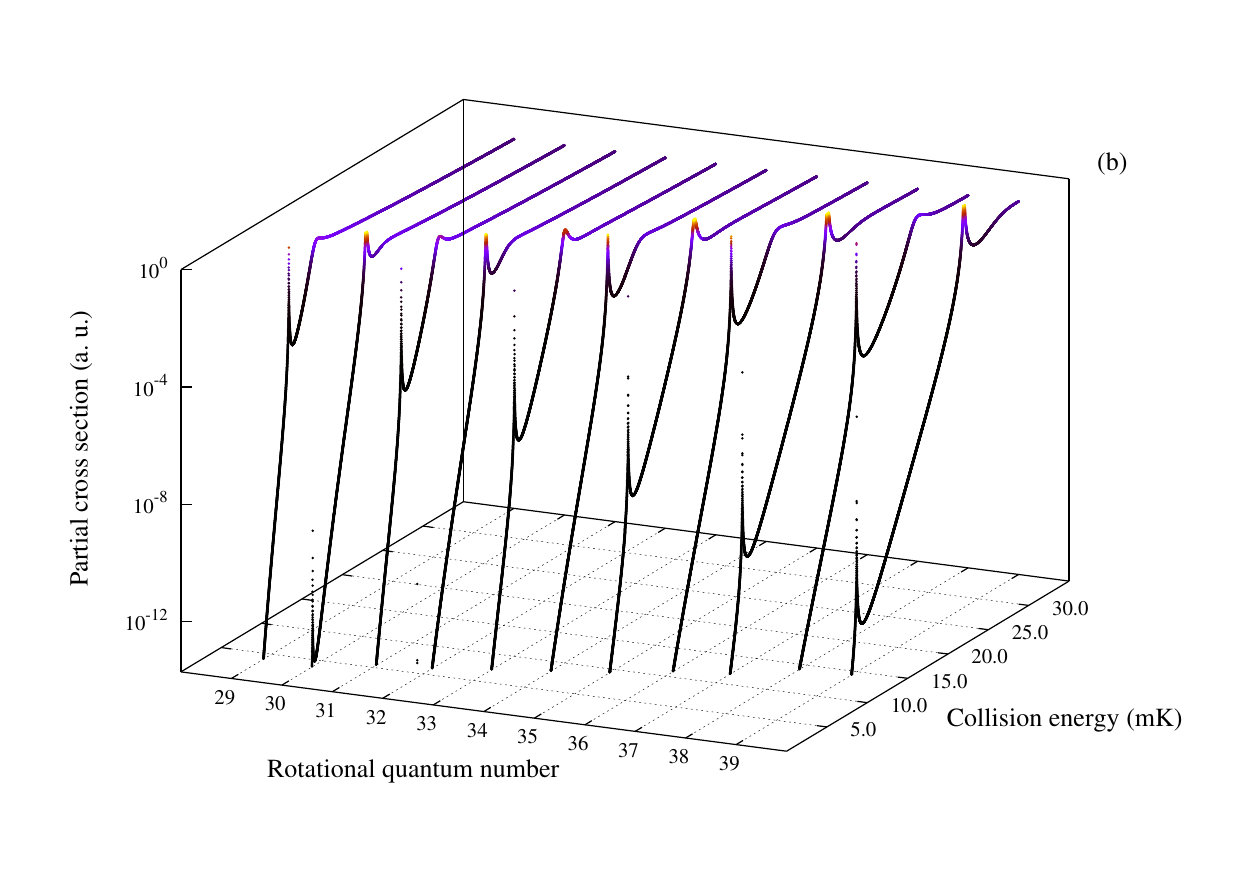}
\caption{(a) Centrifugal barrier for $J=35$ in the A$^1\Sigma^+$ potential energy curve of RbSr$^+$ (dashed line), with radial wave functions vertically placed at the appropriate collision energies, namely 0.08, 0.015, and 0.1~cm$^{-1}$ to illustrate the various tunneling regimes. (b) Partial cross sections as a function of the collision energy (in mK) at 11 given values of rotational quantum numbers ($J'=29$ to 39). Shape resonances gradually evolve with increasing $J$.}
\label{fig:shapewf}
\end{figure}

Anticipating on the results of Section~\ref{sec:rates}, Fig.~\ref{fig:shapewf}a illustrates this well-known feature in the case of RbSr$^+$, where examples of radial wave functions are drawn for three collision energies: (i) at 0.08~cm$^{-1}$, above the centrifugal barrier; (ii) at 0.015~cm$^{-1}$, showing a major tunneling through the barrier, (iii) at 0.01~cm$^{-1}$, with no more tunneling. The last two cases correspond to the Wigner threshold law energy range, and the related partial cross sections reported in Fig.~\ref{fig:shapewf}b present an energy dependence proportional to $k^2$ as expected for high partial wave value associated with a $R^{-4}$ long-range interaction. The wave function of type (ii) above gives rise to sharp shape resonances with widths from a fraction of $\mu$K up to about 4~mK, requiring a very small energy step (2$\mu$K) to locate them. The first case depicts a regime similar to the Langevin one, with the same $k^{-1}$  energy dependence of the cross-section at high energy in every partial wave, but with a magnitude smaller than the one predicted by the Langevin model (see Section~\ref{sec:rates}). Note also the alternation of the occurrence of the shape resonances: if a shape resonance shows up for the partial wave $\ell$, another also occurs for the $\ell+2$ partial wave as demonstrated  by B.~Gao in the case of a $R^{-4}$ long range interaction \cite{gao2013}.

\section{Electronic structure calculations for (Alk-Alke)$^+$ molecular ions}
\label{sec:structure}

The potential energy curves, permanent electric dipole moments (PEDMs) and transition electric dipole moments are computed following the same method described in details in Ref.~\cite{aymar2005,aymar2006a,aymar2011}. We briefly recall here the main steps of the calculations,  carried out using the Configuration Interaction by Perturbation of a Multiconfiguration Wave Function Selected Iteratively (CIPSI) \cite{huron1973} package developed at "Laboratoire de Chimie et Physique Quantiques, Universit$\acute{e}$ Toulouse III - Paul Sabatier" (France). The (Alk) + (Alke)$^{+}$ systems are modeled as molecules with two valence electrons moving in the field of the (Alk)$^+$ and (Alke)$^{2+}$ ions represented with ECPs including relativistic scalar effects. These ECPs are taken from Refs. \cite{durand1974,durand1975} for all species but Sr$^{2+}$ and Ba$^{2+}$ \cite{fuentealba1985,fuentealba1987} and Yb$^{2+}$ \cite{wang1998}. The ECPs are complemented with core polarization potentials (CPPs) depending on the orbital angular momentum $\ell$ of the valence electron of Rb (or Li) and (Alke)$^+$ \cite{muller1984a,foucrault1992}, and parametrized with the Rb$^+$ (or Li$^+$) and (Alke)$^{2+}$ static polarizabilities $\alpha$ and two sets of three cut-off radii $\rho_s$, $\rho_p$, and $\rho_d$. All these parameters (except for Yb$^+$) were determined in our previous works on alkali-atom-strontium neutral molecules \cite{guerout2010} and on alkaline-earth hydride ions \cite{aymar2012}. The value for the Yb$^{2+}$ polarizability is $\alpha=$ 6.388~a.u. \cite{dzuba2010}, and the cut-off radii $\rho_s=$ 1.8869~a.u., $\rho_p=$ 0.89235~a.u., and $\rho_d=$ 2.051150~a.u. have been adjusted to reproduce the experimental energies of the $ f^{14}6s$, $4f^{14}6p$ and $4f^{14}5d$ states of the Yb$^+$ ion \cite{nist_database}.  Only the remaining two valence electrons are used to calculate the Hartree-Fock and the excitation determinants, in an atom-centered Gaussian basis set, through the usual self-consistent field methodology. The basis set used for the Rb and Li atoms are from Refs.~\cite{aymar2005,aymar2006a} and from Ref.~\cite{aymar2012,bouissou2010} for the (Alke)$^+$ ions (except for Yb$^+$). We used for Yb$^+$ a large uncontracted $(5s5p6d)$ Gaussian basis set with the series of exponents (0.785942, 0.370489, 0.068843, 0.031984, 0.015622), (0.685598, 0.338376, 0.073598, 0.034232, 0.015836) and (1.377363, 0.785942, 0.492827, 0.1680780, 0.058334, 0.02) for $s$, $p$, and $d$ orbitals, respectively. It is worthwhile to note that to our knowledge this is the first time that the Yb$^+$ ion is represented in a so-called ```large core''' model, apart from Ref.~\cite{wang1998}. A more detailed analysis of the (Yb-Alk)$^+$ compounds (where Alk is an alkali-metal atom among Li, Na, K, Rb, Cs) will be presented in a forthcoming study.

\begin{figure}[!htbp]
\centering
\includegraphics[scale=0.5, angle=0]{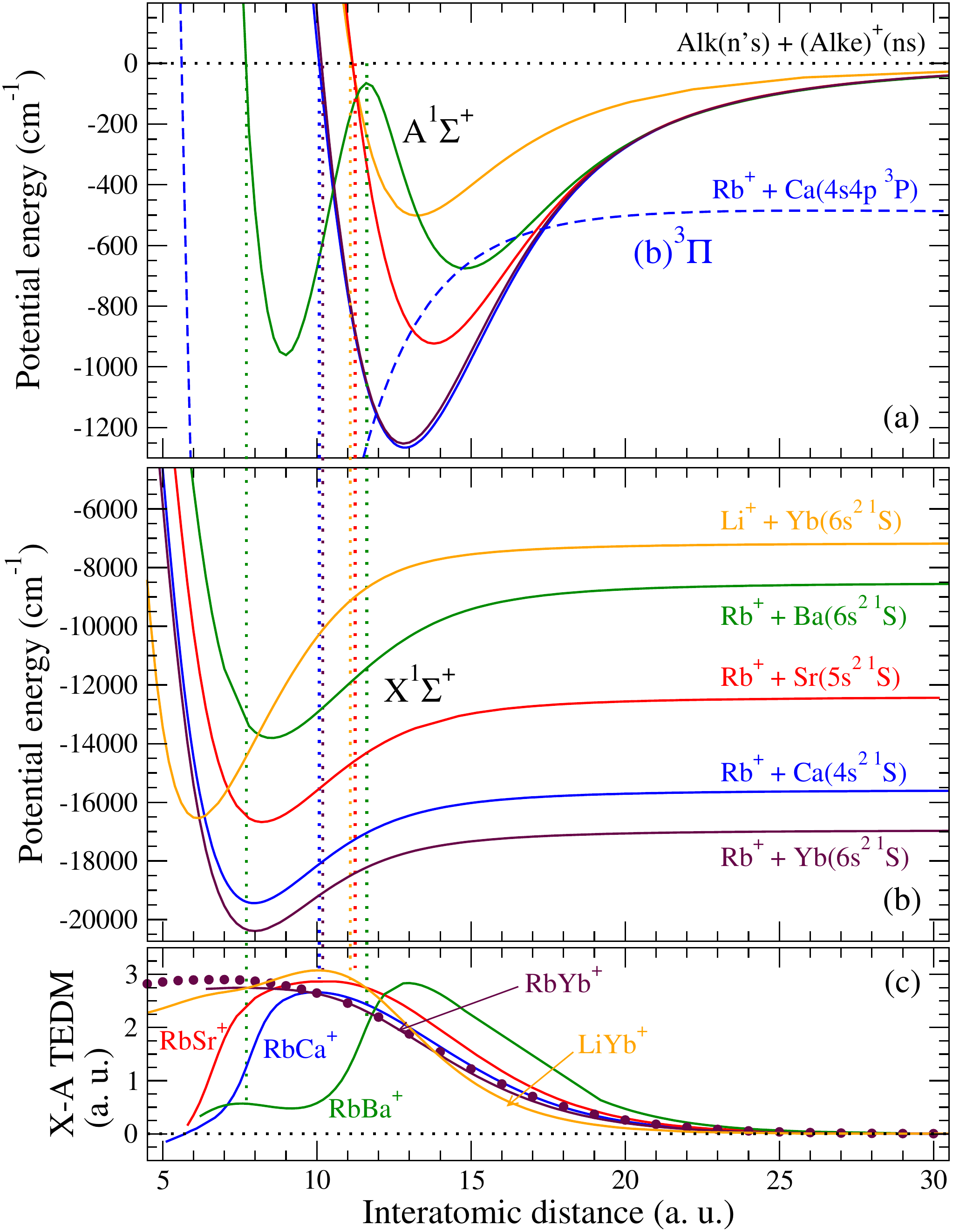}
\caption{Potential energy curves of (a) the A$^1\Sigma^+$ state, and (b) of the X$^1\Sigma^+$ state, for the [(Alk)-(Alke)]$^+$ molecules, with (Alk)=Li or Rb ($n'=2, 5$), and (Alke) = Ca, Sr, Ba, and Yb ($n=$4, 5, 6, 6). The origin of energies is taken at the dissociation of the A$^1\Sigma^+$ PEC for all systems. The lowest $^3\Pi$ PEC of Rb-Ca$^+$ is also displayed for completeness. (c) Transition electric dipole moments between the X and A states. Open circles: Ref.~\cite{mclaughlin2014}. The vertical dotted lines tag the inner turning points of the incoming continuum wave function -- in the entrance channel -- with the respective points in the exit channel and the transition dipole moment magnitude at that position.}
\label{fig:pec}
\end{figure}

A full configuration interaction (FCI) is finally achieved to obtain the relevant PECs and TEDMs displayed in Fig.~\ref{fig:pec}, limited to the X and A states and their transition. The origin of energies is taken at the (Alke)$^+$($ns$) + Rb($5s$) (or Yb$^+$($6s$) + Li($2s$)) dissociation limit corresponding to the entrance channel relevant for the present study. By construction of our approach, the energies of the Rb($5s$), Li($2s$), and Yb$^+$($6s$) ground levels are adjusted to the experimental values. Thus the dissociation energy matches the experimental one in the former case, while for the latter case it results from the CI calculation performed on the neutral (Alke) atom. These calculated (Alke) ground-state ionization energies are reported in Ref.~\cite{aymar2012} for all atoms except Yb. In this latter case, we obtained an ionization energy of 50330~cm$^{-1}$, smaller by 113~cm$^{-1}$ compared to the experimental one (50443~cm$^{-1}$ \cite{nist_database}). This discrepancy of 113~cm$^{-1}$ also holds for the spacing between the Yb$^+$($6s$)+(Alk)($ns$) and the Yb($6s^2\,^1S$)+(Alk)$^+$ asymptotes (equal to 16864.8~cm$^{-1}$ in our method, in good agreement with the experimental value of 16752.2~cm$^{-1}$ \cite{nist_database}). It favorably compares to the discrepancy of 322~cm$^{-1}$ reported in Ref.~\cite{lamb2012}, and to Ref.~\cite{sayfutyarova2013} which reported an energy difference of 16279~cm$^{-1}$ between these two asymptotes. Finally, in their work on LiYb$^+$ \cite{tomza2015} the authors report an ionization energy for the Yb ground state of 50267~cm$^{-1}$.

All PECs for the A state consistently behave in the same way at large distances (Fig.~\ref{fig:pec}a), varying as $-C_4/(2R^4)$ determined by the Rb (or Li) static polarizability $\alpha \equiv 2C_4$. A rough fit with a single parameter of the long-range part of the PECs yields $2C_4(\textrm{Rb}) \approx 360 \pm 20$~a.u. (depending on the Rb(Alke)$^+$ species) and $2C_4(\textrm{Li}) \approx 192$~a.u.. Note that a more relevant determination using our quantum chemistry approach above, in the context of the finite field method, yields $2C_4(\textrm{Rb}) =318$~a.u. and  $2C_4(\textrm{Li})=164$~a.u. \cite{deiglmayr2008} in good agreement with the various determinations of the Rb static polarizability \cite{derevianko2010}. The potential well of this state results from a strongly avoided crossing with the lowest X$^1\Sigma^+$ PEC, which is located at about the same distance range for all species. The well-depth decreases with increasing reduced mass along the (RbAlke)$^+$ series. As also observed in previous works \cite{knecht2008,knecht2010,krych2011}, the RbBa$^+$ PEC exhibits a double well induced by an avoided crossing with an upper $^1\Sigma^+$ curve not displayed here. This feature is imprinted in the variation of the corresponding A-X TEDM (Fig.~\ref{fig:pec}c). Surprisingly, the PEC of RbCa$^+$ and RbYb$^+$ are very similar, which could be related to the well-known $f$-shell contraction in the lanthanide atoms, yielding to the Yb the character of a much smaller atom than expected from its mass \cite{cowan1981}. In contrast, the A potential well in LiYb$^+$ is significantly less deep than for the (RbAlke)$^+$ series due to the low value of the $C_4$ long-range coefficient.

The PECs for the X state all look very similar along the (RbAlke)$^+$ series (Fig.~\ref{fig:pec}b). Due to the smaller size of the Li$^+$ ion compared to the Rb$^+$ one, the minimum of the well for LiYb$^+$ is located at quite short distances, while its depth is larger than for the Rb-Alke$^+$ compounds. As above, the long-range part of the PEC varies as $R^{-4}$, with a coefficient depending on the static polarizability of the (Alke) neutral species. Again, a simple fit of this long-range part with a single parameter gives $2C_4 \approx$ 140~a.u., 220~a.u., 300~ a.u., 154~a.u., and 154~a.u., for RbCa$^+$, RbSr$^+$, RbCa$^+$, RbYb$^+$ and LiYb$^+$ respectively, in reasonable agreement with the static polarizabilities of Ca, Sr, Ba, and Yb from our previous calculations \cite{aymar2012} or from \cite{derevianko2010}. Note again that to determine more rigorously these quantities within our quantum chemistry approach, a method like the finite-field approach should be used, as we have done for instance in the Sr case \cite{aymar2011}.

The main spectroscopic constants of these PECs are deduced from a parabolic fit of the bottom of the wells around their minimum, and are reported in Tables~\ref{tab:specX} and \ref{tab:specA}, allowing for comparison with the few other theoretical determinations available in the literature. The choice of isotopologues is somewhat arbitrary, as it has no significant influence on the present scattering calculations, as it will be discussed later.

There is only one other work reporting about the molecular structure of RbCa$^+$ \cite{tacconi2011,belyaev2012} for the study of NRCT. The authors used a so-called "`small core"' ECPs where electrons from the $4s$ and $4p$ shells in Rb, and from the $3s$ and $3p$ shells in Ca$^+$ are explicitly treated besides the valence electrons. As for the $^1\Sigma^+$ states, the authors only provided figures for the PECs from which we estimated equilibrium distances and potential well depths in good agreement with our results. This is a strong indication in favor of the ability of our approach employing large-core ECPs to represent core-valence electronic correlation.

The RbBa$^+$ molecule already attracted several detailed non-relativistic studies which are in quite good agreement among each other. We observe the same trend than for RbCa$^+$, \textit{i.e.} for both X and A states we find slightly smaller equilibrium distances, and slightly deeper potential wells than in Refs.~\cite{knecht2008,knecht2010,krych2011}. In Ref.~\cite{krych2011} the authors investigated the role of the triple excitations in their approach, yielding a double-well potential for the A state with a barrier located below the dissociation limit at about the same position (estimated from their figure) than in our calculation. The TEDM between the X and A states display a very similar shape with two maxima in both our results and in those of Ref.~\cite{krych2011}, reflecting the double-well structure of the A PEC. The largest value of $d^A_{\textrm{m}}=$2.84~a.u. is reached at a distance of $R^A_{\textrm{m}}=$13~a.u. in our work, compared to $d^A_{\textrm{m}} \approx $2.65~a.u. at $R^A_{\textrm{m}}=$13.5~a.u. in Ref.~\cite{krych2011}. Again, these are strong arguments in favor of the present approach based on ECP and CPP to represent core-valence correlation effects with the same quality than in other methods.

For RbSr$^+$, there is no other calculation available than \cite{aymar2011}, but given the favorable comparisons above, and as this ion is treated consistently with the same method than the three other species, the present results are probably of satisfactory quality.

As already quoted above, the cases of RbYb$^+$ and LiYb$^+$ are peculiar, as the Yb$^+$ ion and Yb atom have not been treated with large ECP before. Tables~\ref{tab:specX} and \ref{tab:specA} reveal a contrasted situation for RbYb$^+$ in comparison with the results of Refs.~\cite{lamb2012,mclaughlin2014,sayfutyarova2013} that are all obtained with similar approaches based on small core ECPs. Our values for $R_e$ and $D_e$ of the X PEC are in good agreement with those of Ref.~\cite{sayfutyarova2013}, while the potential well depth from Refs. \cite{lamb2012,mclaughlin2014} is twice smaller than ours, with a minimum located at a significantly larger distance. This has been quoted by the authors of Ref.~\cite{mclaughlin2014} but no explanation has been provided. On the other hand for the A state, the three previous papers have obtained results in good agreement among each other, while we find an equilibrium distance shorter by about 1~a.u. and the well depth deeper by about 400~cm$^{-1}$. Surprisingly the magnitude and the $R$ variation of the TEDM of the X-A transition look very similar among all determinations including ours, with a maximal value of about 3 a.u. (Fig.~\ref{fig:pec}), suggesting that electronic wave functions behave in the same way in all methods.

The only other work about the LiYb$^+$ structure that we are aware of is the very recent article by Tomza \textit{et al.} \cite{tomza2015}, using the same approach than in Ref.~\cite{krych2011} for RbBa$^+$. The conclusions looks similar to the case of RbYb$^+$ and RbBa$^+$ for both the X and A PECs. The spectroscopic constants of the X PEC are in good agreement, with a smaller harmonic constant and a smaller potential well by about 5\% in our results compared to those of Ref.~\cite{tomza2015}. We found the depth of the A PEC about 30\% larger,  with a minimum located at a distance by about 1.15~a.u., and an harmonic constant smaller by about 3.5\%, than the values obtained in Ref.~\cite{tomza2015}. In contrast with these large differences, the present TEDM between the X and A states are in very good agreement with this latter work, as already observed for the other ions above.

\begin{table}[!htbp]
\caption{Main spectroscopic constants for the X$\sspec$ electronic state of Rb(Alke)$^{+}$ and LiYb$^{+}$: the bond length, $r_{e}$ (a. u.), the harmonic constant, $\omega_{e}$ (cm$^{-1}$), the potential well depths, $D_e$ (cm$^{-1}$), the rotational constant, $B_{e}$ (cm$^{-1}$).  Other published values are also provided when available. Two sets of results were published in Ref.~\cite{knecht2010}, (a) using a multireference configuration interaction (MRCI), and (b) using the coupled-cluster method. The $^*$ symbol for Ref.~\cite{tomza2015} indicates that the authors reported spectroscopic constants for the $^{7}$Li$^{172}$Yb$^+$ isotope.}
\begin{center}
\begin{tabular}{rrrrr}
\hline \hline
(X)$\sspec$            &$r_{e}$     &$\omega_{e}$&$B_{e}$&$D_e$         \\
\hline
$^{87}$Rb$^{40}$Ca$^+$ &            &            &       &              \\
this work              &7.96        &73.02       &0.0346 &3850.9        \\
\cite{tacconi2011}     &$\approx$8.0&     -      &   -   &$\approx$3730 \\
$^{87}$Rb$^{87}$Sr$^+$ &            &            &       &              \\
this work              &8.23        &59.41       &0.0205 &4265.7        \\
\cite{aymar2011}       &8.2         &58          &   -   &4285          \\
$^{87}$Rb$^{137}$Ba$^+$&            &            &       &              \\
this work              &8.53        &51.01       &0.0155 &5292.5        \\
\cite{knecht2010}$^a$  &8.72        &51.77       &   -   &5055.0        \\
\cite{knecht2010}$^b$  &8.75        &52.79       &   -   &5034.0        \\
\cite{krych2011}       &8.67        &     -      &   -   &5136          \\
$^{87}$Rb$^{173}$Yb$^+$&            &            &       &              \\
this work              &7.99        &49.73       &0.0162 &3435.88       \\
\cite{sayfutyarova2013}&8.08        &    -       &   -   &3496          \\
\cite{lamb2012}        &9.03        &33.77       &0.0127 &1776          \\
\cite{mclaughlin2014}  &9.0         &    -       &   -   &1807.4        \\
$^{7}$Li$^{173}$Yb$^+$ &            &            &       &              \\
this work              &6.14        &223.4       &0.2370 &9383.7        \\
\cite{tomza2015}$^*$   &6.20        &231         &0.23   &9412          \\
\hline \hline
\end{tabular}
\end{center}
\label{tab:specX}
\end{table}

\begin{table}[!htbp]
\caption{Same as Table~\ref{tab:specX} for the A$\sspec$ electronic states of Rb(Alke)$^{+}$ and LiYb$^{+}$. In addition the transition energy $T_e$ (cm$^{-1}$) between the bottom of the A and X potential wells is reported. For RbBa$^+$, values for the inner and outer wells are displayed on the same line. The position of the top of the barrier, and its energy below the dissociation limit, are also indicated under the $r_e$ and $D_e$ columns. The $^*$ symbol for Ref.~\cite{tomza2015} indicates that the authors reported spectroscopic constants for the $^{7}$Li$^{172}$Yb$^+$ isotope.}
\begin{center}
\begin{tabular}{rrrrrr}
\hline \hline
(A)$\sspec$            &$r_{e}$     &$\omega_{e}$&$B_{e}$&$D_e$   &$T_{e}$  \\
\hline
$^{87}$Rb$^{40}$Ca$^+$ &            &            &       &             &       \\
this work              &12.86       &29.89       &0.0133 &1271.64      &18171.2 \\
\cite{tacconi2011}     &$\approx$13 &     -      &   -   &$\approx$1170&  -     \\
$^{87}$Rb$^{87}$Sr$^+$ &            &            &       &             &        \\
this work              &13.79       &20.94       &0.0073 &933.0        &15752.3 \\
\cite{aymar2011}       &13.8        &21          &   -   &960          &  -     \\
$^{87}$Rb$^{137}$Ba$^+$&            &            &       &             &        \\
this work              &8.98/14.81  &40.49/15.94 &0.0140/0.0051&964.81/679.27&12849.8/13135.3  \\
(inner/outer)          &            &            &       &             &        \\
this work              &11.61       &     -      &   -   &65.3         &   -    \\
(barrier)              &            &            &       &             &        \\
\cite{knecht2010}$^a$  &9.03        &     -      &   -   &  -          &   -    \\
\cite{knecht2010}$^b$  &            &            &   -   &             &        \\
\cite{krych2011}       &9.02/15.19  &      -     &   -   &911/576      & 12569/12904       \\
(inner/outer)          &            &            &       &             &        \\
$^{87}$Rb$^{173}$Yb$^+$&            &            &       &             &        \\
this work              &12.82       &21.08       &0.0063 &1258.25      &19138.2 \\
\cite{sayfutyarova2013}&13.8        &    -       &   -   &836.0        &   -    \\
\cite{lamb2012}        &14.362      &16.807      &0.00505&875.1        &   -    \\
\cite{mclaughlin2014}  &14.0        &    -       &   -   &875.1        &   -    \\
$^{7}$Li$^{173}$Yb$^+$ &            &            &       &             &\\
this work              &13.25       &46.57       &0.0509 &501.7        &16046 \\
\cite{tomza2015}$^*$   &14.4        &37.1        &0.045  &358          &15857 \\
\hline \hline
\end{tabular}
\end{center}
\label{tab:specA}
\end{table}

\begin{figure}[!htbp]
\centering
\includegraphics[scale=0.6, angle=0]{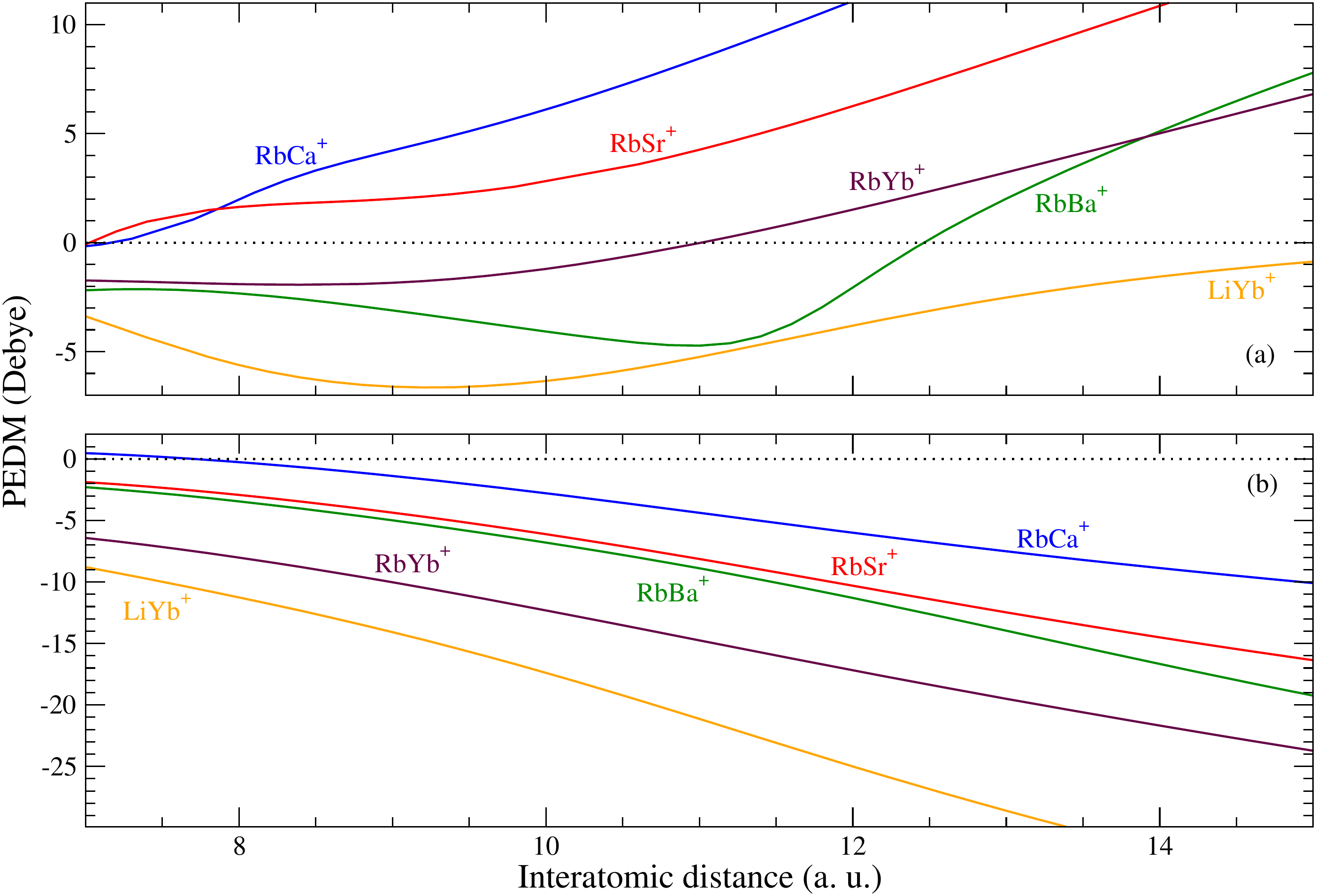}
\caption{Permanent electric dipole moments of (a) the A$^1\Sigma^+$ state, and (b) of the X$^1\Sigma^+$ state, for the Rb(Alke)$^+$ molecules, with (Alke) = Ca, Sr, Ba, and Yb, and for Li-Yb$^+$.}
\label{fig:pedm}
\end{figure}

For completeness we report in Fig.~\ref{fig:pedm} the PEDMs for both the A and X states of these molecular ions. They have been computed with respect to the origin of coordinates placed at the center-of-mass of the molecule, with the axis oriented from the Rb or Li atom towards the (Alke)$^+$ ion. We note the total mass $M=M(\textrm{(Alk)}^+)+M(\textrm{Alke})$ and the mass difference $\delta M= M(\textrm{(Alk)}^+)-M(\textrm{Alke})$. We checked that at large distances - beyond 20~a.u., typically- the PEDMs linearly diverge as $R/2 (1+\delta M /M)$ and $-R/2(1-\delta M /M)$ for the A and X states respectively, as the charge is carried either by the (Alke) species or by the (Alk) one. The PEDM of the A state in RbBa$^+$ has an abrupt change of slope around 12~a.u. due to the occurrence of the barrier between the two potential wells. Tomza \textit{et al.} \cite{tomza2015} have displayed the absolute value of the X and A PEDMs in LiYb$^+$, showing very similar variation to ours but with a slightly smaller amplitude.

The data for PECs, TEDMs, and PEDMs are collected in the Supplemental Material attached to this paper \cite{supp-mat}.

\section{Cross sections and rates for RA and RCT}
\label{sec:rates}

The main objective of this study is to evaluate, through the comparison of five molecular ions treated consistently by the same approaches for both their electronic structure and their dynamics, the efficiency of the formation of molecular ions, and the possibility to observe shape resonances with the current experimental resolution. We performed about 40000 calculations per system in the collisional energy range between 100~$\mu$K to 80~mK, for all the total rotational quantum numbers between 0 and 80. The energy step was fixed to 2$\mu$K in order to properly locate all the shape resonances.

\begin{figure}[!htbp]
\centering
\includegraphics[scale=0.5, angle=0]{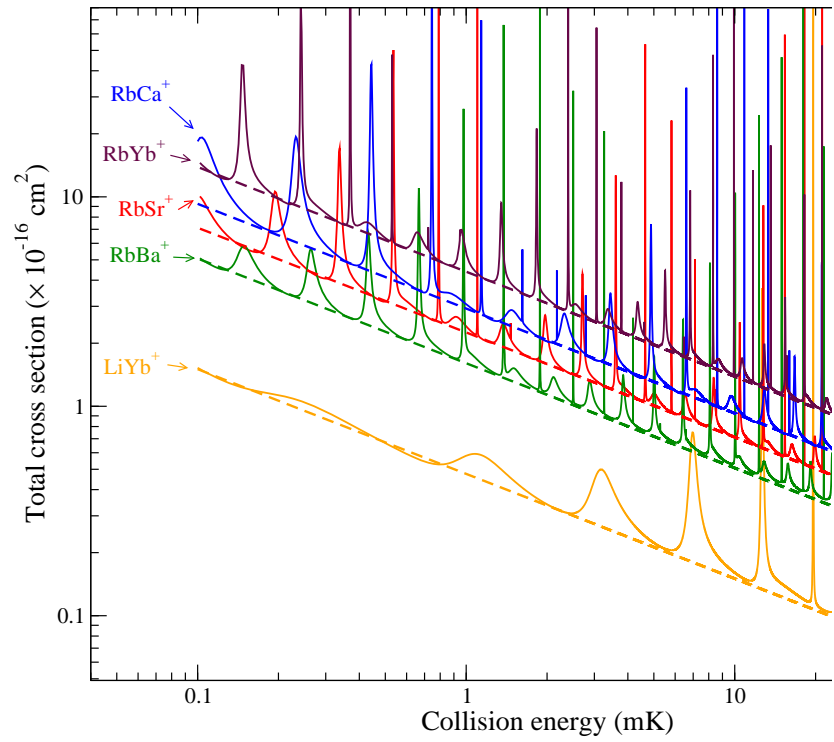}
\caption{Calculated total cross sections (RA + RCT) from Eqs.~(\ref{eq:csrct}) and (\ref{eq:csra}), as functions of the collision energy (with both axes scaling logarithmically). The dashed lines correspond to the classical cross section of Eq~(\ref{eq:langevin}).}
\label{fig:total_cs}
\end{figure}

Figure~\ref{fig:total_cs} displays the total cross sections (RA + RCT) for all five systems, which present several similar features. As anticipated in section \ref{sec:theory} they all present the same  $\epsilon_i^{-1/2}$ dependence with the entrance channel energy $\epsilon_i$ which corresponds to the Langevin-type regime. However the magnitude of the cross sections is much smaller than the one predicted by the original Langevin model \cite{langevin1905}
\begin{equation}
\sigma_{\textrm{Lang}}= p P_{\textrm{Lang}} \pi \epsilon_i^{-1/2} \sqrt{2C_4}
\label{eq:langevin}
\end{equation}
with the statistical factor $p=1/4$ and a reaction probability $P_{\textrm{Lang}}$ following the capture equal to 1. As the baseline of our calculated cross sections (\textit{i.e.} ignoring shape resonances) varies as $\epsilon_i^{-1/2}$ (Fig~\ref{fig:total_cs}), we find a constant probability $P_{\textrm{Lang}}=0.88\times 10^{-5}$, $2.2 \times 10^{-5}$, $3.0\times 10^{-5}$, $3.9\times 10^{-5}$, and $5.9\times 10^{-5}$ for LiYb$^+$, RbBa$^+$, RbSr$^+$, RbCa$^+$, and RbYb$^+$, respectively. 

Numerous narrow shape resonances arising from the entrance channel can be observed in each system, with a density increasing with the reduced mass. In this respect, only a few resonances are predicted in LiYb$^+$ due to its comparatively small reduced mass. The resonances have obviously energy positions which dramatically depend on the calculated potential. By comparing the baseline above to a $\epsilon_i^{-1/2}$ fit of the total cross-sections, we estimate the contribution of the shape resonances to about 20$\pm 2$ \% for all these species over the displayed energy range. Apart from the resonances, the hierarchy of the cross sections reflected by their baselines mostly reflects the increasing magnitude of the $\omega_{if}^3$ factor in Eqs.~(\ref{eq:csrct}) and (\ref{eq:csra}) along the series LiYb$^+$, RbBa$^+$, RbCa$^+$, RbSr$^+$, and RbYb$^+$ which can be roughly appreciated from the relative position in energy of the relevant turning points in the X and A PECs (Fig.~ \ref{fig:pec}). 

\begin{figure}[!htbp]
\centering
\includegraphics[scale=0.5, angle=0]{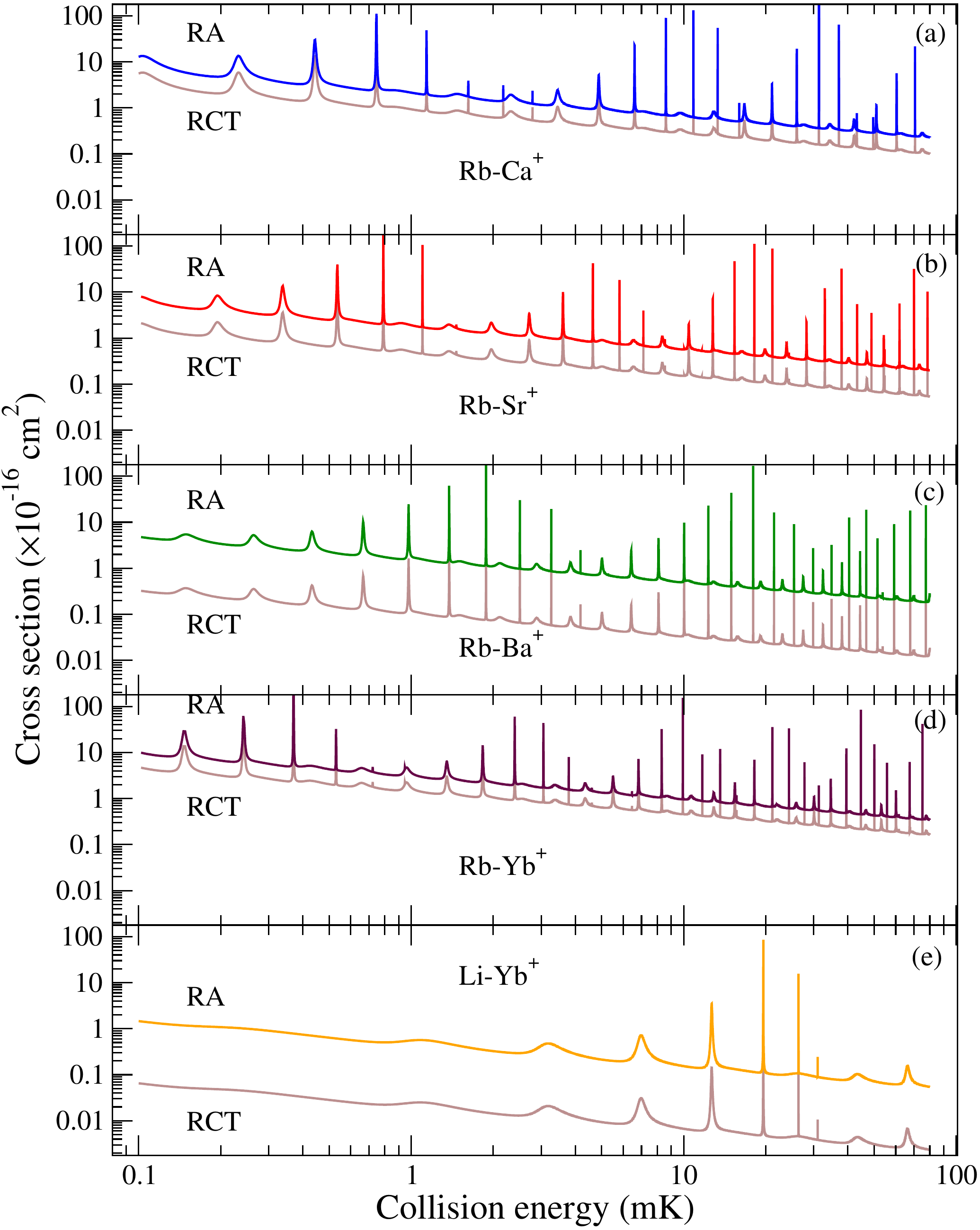}
\caption{Calculated total cross sections as functions of the collision energy (with both axes scaling logarithmically), for radiative association (RA) and radiative charge transfer (RCT) (see Eqs.~(\ref{eq:csrct}) and (\ref{eq:csra})).}
\label{fig:total_cs_RA_RCT}
\end{figure}

The RA process is predicted to dominate for all systems in comparison with RCT, by a factor $\approx 2.4$ for RbCa$^+$, $\approx 3.8$ for RbSr$^+$, $\approx 13.6$ for RbBa$^+$, $\approx 2.1$ for RbYb$^+$, and $\approx 22.4$ for LiYb$^+$ (Fig.~\ref{fig:total_cs_RA_RCT}). This is due to the favorable relative position of the X and A potential wells which are shifted against each other in $R$, so that the classical inner turning point of the A potential is aligned with the one of quite deeply bound levels of the X potential (see dotted lines in Fig.~\ref{fig:pec}). The very similar behavior between RbCa$^+$ and RbYb$^+$ reflects the strong similarities of the PECs and TEDMs functions. The RA process is even more favored in RbBa$^+$ due to the double well in the A potential curve, which enhances the spontaneous emission towards X bound levels around the top of the barrier where the TEDM is large. In all cases, the shape resonances in the RA and RCT cross sections are the same, as they are due to the centrifugal barrier in the entrance channel (see also Refs. \cite{belyaev2012,sayfutyarova2013}). In the framework of the present model, the choice of the isotopologues for each ionic species does not yield significantly different cross sections in magnitude, as the reduced mass (which differs by less than one percent among a series of isotopologues of a given ion) is involved only through the kinetics described by the radial wave functions. Sayfutyarova \textit{et al.} reached the same conclusion for isotopologues of RbYb$^+$ \cite{sayfutyarova2013}, as well as Tomza \textit{et al.} for LiYb$^+$ \cite{tomza2015}. But of course, the energy positions of the shape resonances will vary with the chosen isotopologue.

\begin{figure}[!htbp]
\centering
\includegraphics[scale=0.5, angle=270]{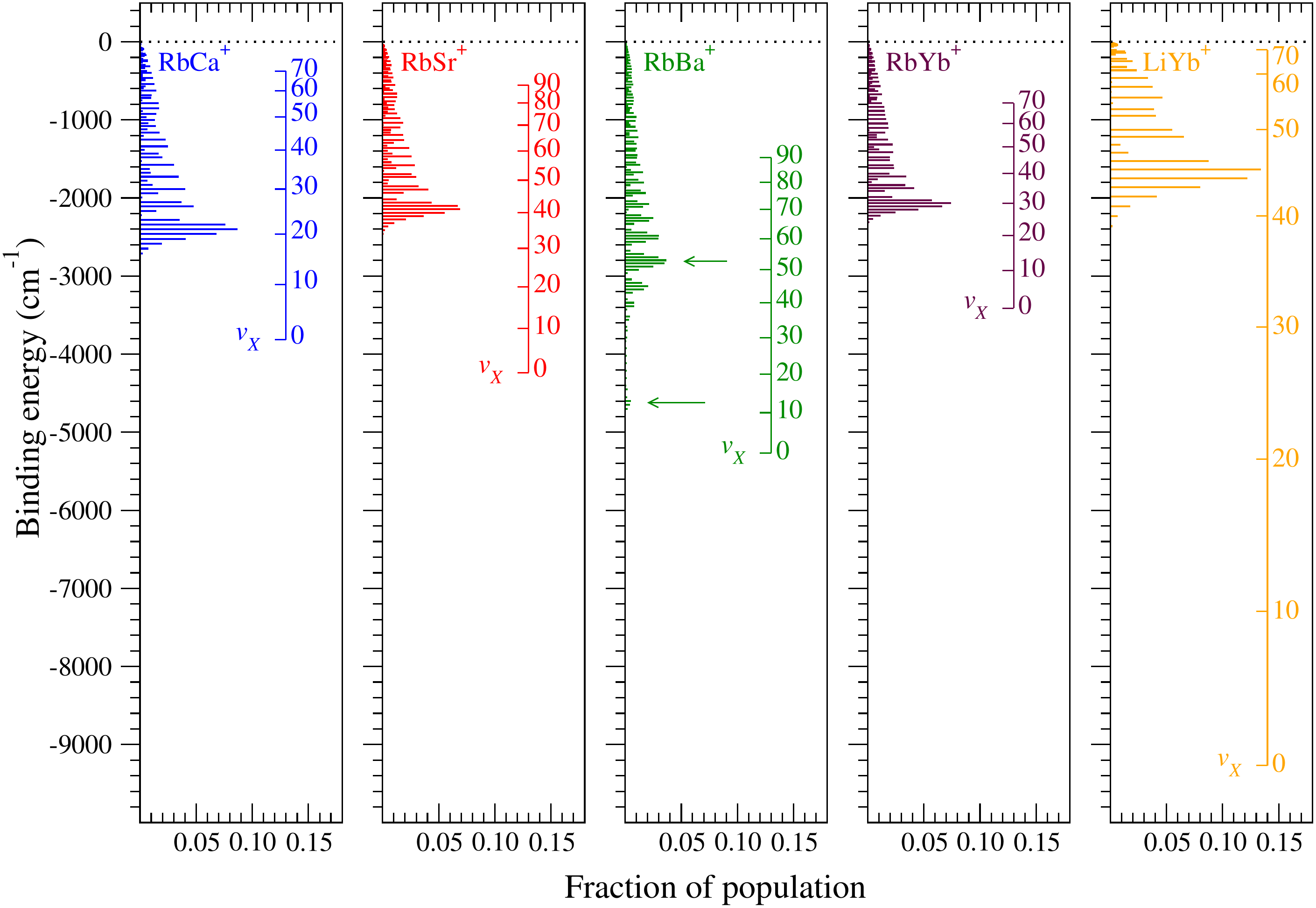}
\caption{Calculated vibrational distributions of the ground state molecular ions produced by radiative association, as a function of the binding energies and the corresponding vibrational quantum numbers, $v_X$, for the rotational quantum number $J = 1$, at $\epsilon_i = $0.1~mK. Note that the vibrational distribution is the same over the displayed energy range. For comparison purpose, the energy scale is common for all species, and thus vibrational indexes are shifted from one species to the other.}
\label{fig:vib_dist}
\end{figure}

In Fig.~\ref{fig:vib_dist} are presented the vibrational distributions of the ground state vibrational levels $v_X$ produced by RA for all species, in the $J=1$ case and $\epsilon_i=0.1$~mK as a representative example. The fraction of population of a given vibrational level $v_X$ of the X state is defined as the ratio of the matrix element in Eq.~(\ref{eq:csra}) for $J'=1$ and $v=v_X$, divided by the sum over all vibrational levels of the X state, for $J'=1$, of the matrix elements. As expected, these distributions are similar for the three ions RbCa$^+$, RbSr$^+$, and  RbYb$^+$, due to their similar electronic structure. The peak in the distributions fulfills the Franck-Condon principle as it is located at a transition energy close to the energy difference between the inner classical turning point of the A potential (located in the 10-12 a.u. range) and the outer turning point of the X potential (see vertical dotted lines in Fig.~\ref{fig:pec}). In contrast, the double-well pattern in RbBa$^+$ is manifested by a distribution extended to much lower vibrational levels than for the other species, due to the inner turning point of the A potential located around 8a.u.. The vibrational distribution is double-peaked as well, the peak at low $v_X$ (resp. high $v_X$) corresponding to the energy difference between the inner classical turning point of the A potential and the inner (resp. outer) turning point of the X potential. The peak at low $v_X$ is small however, due to the small magnitude of the TEDM in the corresponding distance range (see Fig.~\ref{fig:pec}c). Finally, the low reduced mass of LiYb$^+$ is reflected in the low density of vibrational levels in the X state compared to the species above. But the trend is similar, \textit{i.e.} only the uppermost levels are expected to be populated by RA. Note that the overall shape of the vibrational distribution looks the same than in Ref.\cite{tomza2015}, with slightly shifted extrema due to the obtained differences in the potential well depths.

The molecular ions are all expected vibrationally hot, as well as rotationally hot given the number of partial waves involved in the present energy range, even around 1~mK where many partial waves are still involved. Due to their non-vanishing PEDM (Fig.~\ref{fig:pedm}) they can radiatively decay in principle down two lower vibrational levels. We computed the radiative lifetime of a given level $v$ for decaying down to the level $v-1$ for all systems, and found that it always exceeds 10~s because of the slow $R$ variation of the PEDM.

\begin{figure}[!htbp]
\centering
\includegraphics[scale=0.5, angle=0]{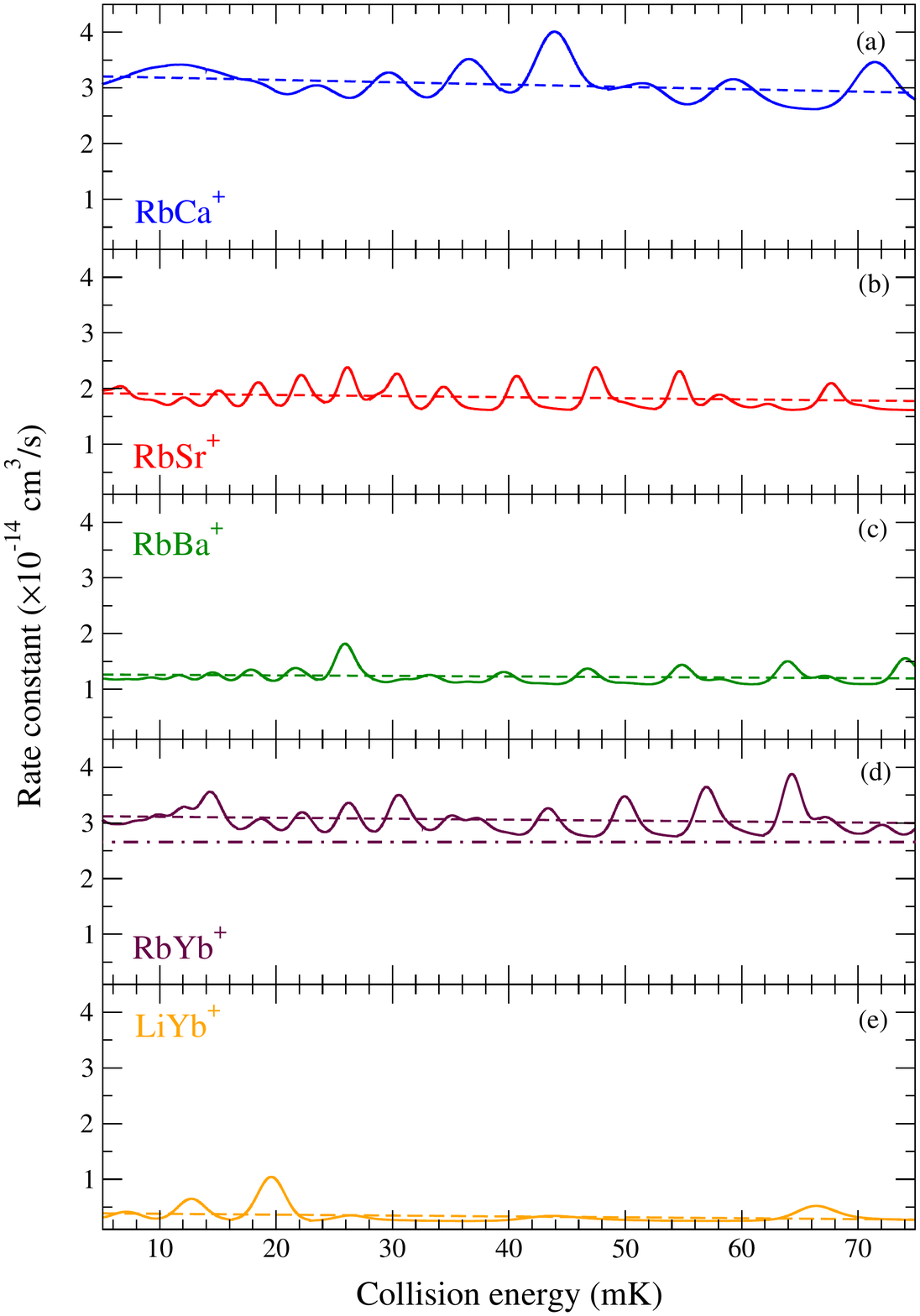}
\caption{Calculated total (RA + RCT) rate constants convoluted from the cross sections of Fig.~\ref{fig:total_cs} with a Gaussian energy distribution of 2~mK half-width. A linear fit of this curves yields an estimate of the average magnitude of the rates (dashed lines). For RbYb$^+$, the calculated rate obtained in Ref.~\cite{sayfutyarova2013,mclaughlin2014} is indicated with a dot-dashed line.}
\label{fig:rates}
\end{figure}

\begin{figure}[!htbp]
\centering
\includegraphics[scale=0.5, angle=0]{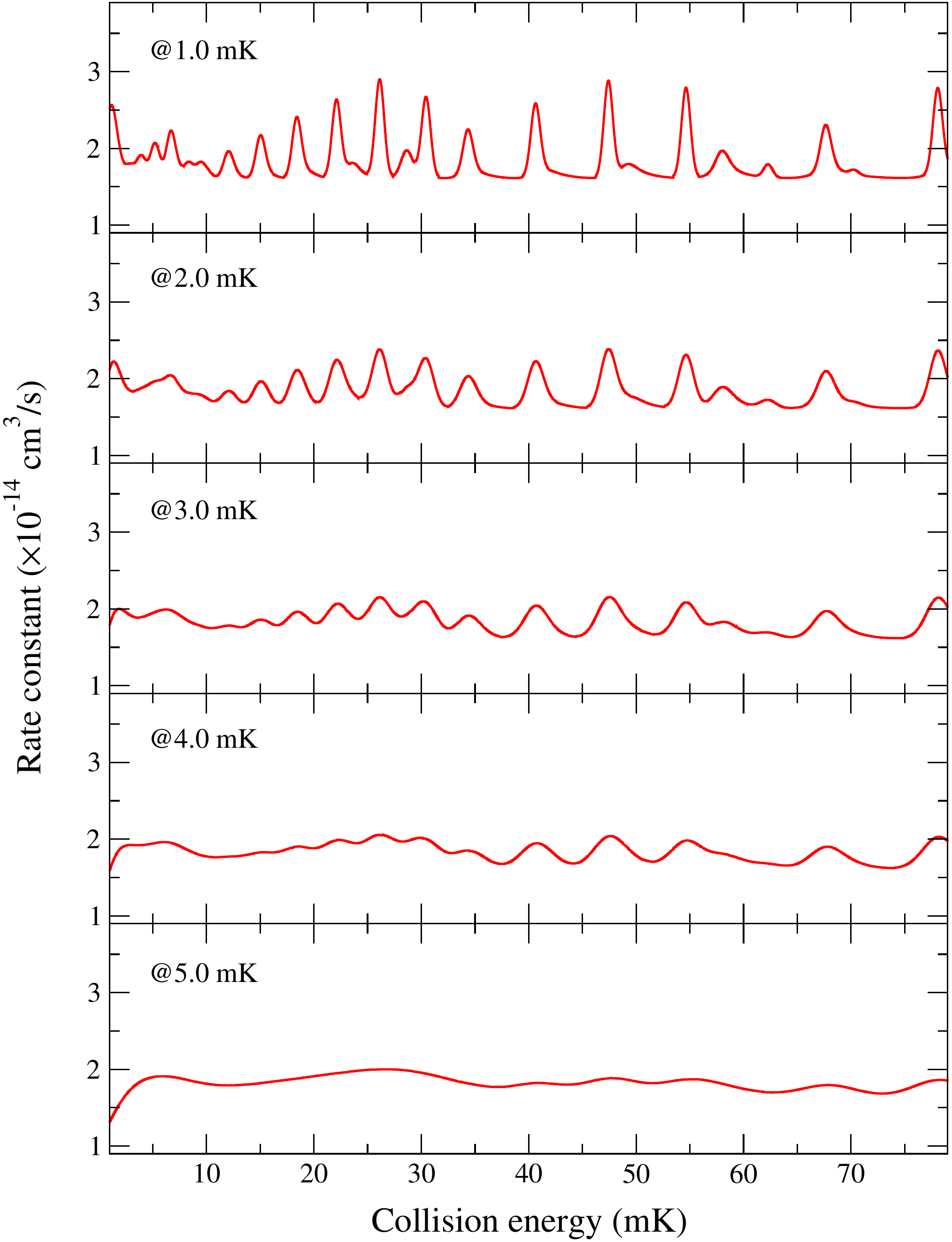}
\caption{Calculated total rate constants for RbSr$^+$ convoluted with a Gaussian distribution half-width from 1~mK to 5~mK.}
\label{fig:gaussian}
\end{figure}

In order to mimic the measured rate constant behavior as a function of the collision energy, the total cross sections (RA + RCT) above must be convoluted with a reliable energy distribution. Such distributions have been modeled using molecular dynamics simulations (see Refs. \cite{hall2011,hall2013a,hall2013b}), reflecting the dominant role of the micromotion of the trapped ions which is much faster than the thermal motion of the ultracold atoms. These energy distributions reflect both the fact that ions are located at well-defined places in the ion trap, and that their micromotion strongly depends on the ion number and thus on their places. This results into quite broad distributions which completely smooth out the shape resonances. Nevertheless, new experimental developments are in progress in various experimental groups aiming at better controlling the relative velocity of the ion and the atom. In order to illustrate the required width which would allow to observe the predicted shape resonances, we convolved the RbSr$^+$ computed rates with Gaussian distributions of half-width varying from 1~mK to 5~mK (Fig.~\ref{fig:gaussian}). We see that a width of about 2~mK would be acceptable. We extended this result to all studied species (Fig.~\ref{fig:rates}), leading to the same statement. As expected from the $\epsilon_i^{-1/2}$ variation of the cross sections, the obtained rates are almost constant over the 1~mK--80~mK collision energy range (see the dashed lines in Fig.~\ref{fig:rates}). Their relative magnitude still reflects the change in the $\omega_{if}^3$ factor invoked earlier for the cross section. For instance, this factor is larger by a magnitude of almost 8 in RbYb$^+$ compared to LiYb$^+$, which can be retrieved in the average level of the rates. It is worthwhile to note that Tomza \textit{et al.} \cite{tomza2015} reported a thermally averaged RA rate almost 100 times larger than the RCT one around 10~mK, and a total rate of about $1.5\times 10^{-14}$~cm$^3$s$^{-1}$, namely about 4 times larger than ours. This latter issue may be due to the statistical factor of 1/4 for populating the singlet state in the entrance channel, which is not mentioned in Ref.\cite{tomza2015}.

\section{Discussion and prospects}
\label{sec:discussion}

We have computed in a consistent way the electronic structure of a series of molecular ions involved in ongoing hybrid experiment merging cold ion and cold atom traps, as well as their radiative emission rate leading either to the formation of cold molecular ions, or to charge exchange. For the first time, molecular ions involving Yb$^+$ or Yb have been modeled using effective large-core potentials completed with a core polarization potential, in the framework of a full configuration interaction approach. Such a simple description provide results for the electronic structure which are generally in good agreement with other determinations, when available. Results for radiative emission rate are computed for the first time for the ionic species RbSr$^+$ mixing an ion and an atom of close mass. The consistent calculation of rates for a series of molecular ions allows us to infer a uniform quality and accuracy over the entire series, which is of great help for comparison with experimental results, as discussed below.

In the experiments quoted in the introduction, the rate measurements are based on the observation of the decreasing of the number of trapped laser-cooled ions with time, due to all possible processes. In some cases, a careful analysis of the data allows for extracting a rate for a particular process. Before discussing the comparison with the published experimental results, it is first worthwhile to check the consistency among all theoretical studies focusing on the radiative processes (RA and RCT) occurring during the collision between two ground state particles.

The dashed lines in Fig.~\ref{fig:rates} indicates the average level of the calculated total radiative rates (RA + RCT), yielding 3.1$\times 10^{-14}$~cm$^3$/s, 1.9$\times 10^{-14}$~cm$^3$/s, 1.2$\times 10^{-14}$~cm$^3$/s, 3.1$\times 10^{-14}$~cm$^3$/s, and 0.31$\times 10^{-14}$~cm$^3$/s, for RbCa$^+$, RbSr$^+$, RbBa$^+$, RbYb$^+$,  and LiYb$^+$, respectively. The remarkably close values of these rates along the species is well understood from the similarities of their electronic structure invoked earlier, while their differences play only a minor role. This is exemplified by the results from Refs. \cite{sayfutyarova2013,mclaughlin2014} of 2.9$\times 10^{-14}$~cm$^3$/s for the RA + RCT rate in RbYb$^+$, where the large difference of the A potential well compared to the present one only slightly influences the final rate, as the main contribution is controlled by the inner classical turning point of the PEC. The magnitude of the present rates is also consistent with the smaller theoretical rate of 1.5$\times 10^{-15}$~cm$^3$/s computed for the same processes in Ca-Yb$^+$ mixture \cite{zygelman2014}, as the energy difference $\delta R^t_{X-A}$ between the inner classical turning point of the A$^2\Sigma^+$ potential curve in the entrance channel, and the outer classical turning point of the X$^2\Sigma^+$ potential curve in the exit channel is smaller (about 9000~cm$^{-1}$) than in the present series of species. Similarly, Makarov \textit{et al.} modeled the same processes in Na-Ca$^+$, and obtained a total rate of 2.3$\times 10^{-16}$~cm$^3$/s which reflects the relatively small difference $\delta R^t_{X-A} \approx 10700$~cm$^{-1}$ (similar to the situation of RbBa$^+$ and LiYb$^+$), combined with a significantly smaller TEDM of about 0.6~a.u. compared to the present cases.

As visible from Fig.~\ref{fig:pec}a the case of RbCa$^+$ is peculiar, due to the presence of the b$^3\Pi$ state dissociating to the Rb$^+$+Ca($4s4p\, ^3P$) limit which is energetically open to the entrance channel. Tacconi \textit{et al.} \cite{tacconi2011} modeled the non-adiabatic coupling related to the spin-orbit interaction between the b$^3\Pi$ state and the A$^1\Sigma^+$ state, thus inducing non-radiative charge transfer (NRCT) which is predicted to dominate the dynamics of the charge exchange process. We note that this process cannot lead to subsequent radiative formation of molecular ions due to spin selection rule. In a further work \cite{belyaev2012}, the NRCT cross sections are found to approximately vary as expected by the Langevin model as $\epsilon_i^{-1/2}$, and to exhibit shape resonances of the same kind than those expected in the RA and RCT cross sections, due to the centrifugal barrier in the entrance channel. Over the present energy range, the NRCT cross section, is evaluated to be about 200 times larger than the total radiative cross section \cite{hall2011}. Around 2~K the NRCT rate is found of the same order of magnitude (3.5$\times 10^{-12}$~cm$^3$/s) than the experimental rate, conservatively estimated of the order of 2 to 3$\times 10^{-12}$~cm$^3$/s \cite{hall2011,hall2013a}. The hypothesis for NRCT has been also invoked in the Ca-Yb$^+$ system, with a rate evaluated to a few $10^{-14}$~cm$^3$/s \cite{rellegert2011}.

As for the Rb-Yb$^+$ system, the experimental rates measured in Ref.~\cite{zipkes2010} are 3$\pm 1 \times 10^{-14}$~cm$^3$/s and 4.5$\pm 1.5 \times 10^{-14}$~cm$^3$/s, corresponding to the $^{172}$Yb$^+$ and the $^{174}$Yb$^+$ isotope, respectively, and when Rb is prepared in the ($F=2, M_F=2$) hyperfine level ($F$ being the total angular momentum including nuclear spin and $M_F$ its projection on the magnetic field axis used for the preparation). The present computed rate as well as those of Refs. \cite{sayfutyarova2013,mclaughlin2014} are thus in remarkable agreement within the error bars with the experimental results. It is noteworthy to quote that the same experimental group later discovered a huge effect of the initial preparation of the Rb atoms in a given hyperfine level, with a rate 35 times larger than the above one when Rb is in the ($F=1, M_F=1$) hyperfine level \cite{ratschbacher2012}. This effect cannot be retrieved at the present level of the theoretical model which do not include the still unexplored dynamics of the numerous hyperfine channels. Moreover the present agreement between experiment and theory suggests that our computed values for LiYb$^+$ rates are correct as well.

The situation for the Rb-Ba$^+$ system is somewhat intermediate in comparison with the two previous cases above. Indeed, there is no possibility for NRCT, and only an upper bound of 5$\times 10^{-13}$~cm$^3$/s is given for the experimental rate in Ref.~\cite{hall2013b}, compatible with the present calculation but suggesting that it is at most 40 times larger than the calculated one. In another experiment with Rb atoms prepared in the ($F=1, M_F=-1$) state in an almost Bose-condensed sample, Schmid \textit{et al.} \cite{schmid2010} estimated an inelastic cross section presumably due to radiative processes between ground state particles in the range of $10^{-15}$~cm$^2$ to $10^{-14}$~cm$^2$ at about 30~mK, corresponding to rate 3 to 30 times larger than the calculated one.

To be complete with such comparisons with experimental results, it is worthwhile to mention the exceptional case of Ca-Yb$^+$ -- not interpreted yet -- where a considerable rate for inelastic collision between ground state particles of $\approx 2\times 10^{-10}$~cm$^3$/s, \textit{i.e.} of the same order of magnitude than quasi-resonant charge transfer reaction between an atom and an ion of the same species but different isotopes (Grier \textit{et al.} \cite{grier2009} measured a rate of $\approx 6\times 10^{-10}$~cm$^3$/s in Yb-Yb$^+$; see also Refs. \cite{cote2000a,lee2013}).

Last but not least, beside this quite contrasted situation for the comparison between experimental rate for inelastic collisions and calculated rates for radiative processes, there is another puzzling and not yet understood feature in this kind of experiments: while the theoretical models all consistently predict that RA should dominate RCT, molecular ions are directly observed by mass spectrometry only in the experiments of Hall \textit{et al.} on Rb-Ca$^+$ \cite{hall2011,hall2013a} and Rb-Ba$^+$ \cite{hall2013b}. An indirect detection of CaYb$^+$ has been also reported in Ref.~\cite{rellegert2011}. This statement strongly suggests that the dynamic in such merged cold atom-cold ion traps is much more complicated than anticipated. Several elements may have a significant impact on the interpretation of the experimental results:
\begin{itemize}
\item The laser light used for cooling and trapping atoms and ions is recognized to have a strong influence on the measured rates, as inelastic processes with excited species may be very large (see for instance Refs.~\cite{hall2011,hall2013a,hall2013b,sullivan2012}). Furthermore, such --quite intense-- light could prevent the produced molecular ions to be detected before being photodissociated. On the theoretical side, such light-assisted processes may be tedious to describe accurately, as much more collision channels are open due to the presence of numerous excited electronic states (see for instance Ref.~\cite{hall2011}) which are more sensitive to the quality of quantum chemistry calculations.
\item The quoted experiment are achieved with various conditions of atomic densities: Rb atoms are either trapped in MOT with a quite low density of a few $10^{9}$atoms/cm$^3$, or in almost Bose-condensed gases with density up to $10^{12}$atoms/cm$^3$. Therefore collisions between surrounding Rb atoms and the formed molecular ions could well occur, and they are strongly exoergic for all systems studied here (see Table~\ref{tab:reactions}), assuming that there is no activation potential barrier in the triatomic collisional complex which would prevent the reaction to occur. One product would be Rb$_2^+$ ions which are indeed observed in the Rb-Ca$^+$ experiment of Refs. \cite{hall2011,hall2013a}, and presumably present in the Rb-Ba$^+$ experiment of Ref.~\cite{hall2013b}. Note that the energy of a Rb$^+$ - Rb(Alke) complex at zero collisional energy is higher than the one of a Rb -Rb(Alke)$^+$ complex (with respect to their common dissociation limit (Rb$^+$+Rb+Ca)) due to the weak binding energy $D_e$ of the Rb(Alke) molecule compared to the one of the related Rb(Alke)$^+$ molecule. Therefore charge exchange reactions between a Rb(Alke)$^+$ ion and a Rb atom is energetically forbidden. 
\item At high atomic density, three-body collisions could well occur, as it has been recently probed \cite{haerter2012,haerter2013}. Such processes may contribute to the formation of molecular ions, as well as to their destruction if they are radiatively created. Experiments of Ref.~\cite{schmid2010} on Rb-Ba$^+$ and of Ref.~\cite{zipkes2010} on Rb-Yb$^+$ both use an almost Bose-condensed atomic sample and have not observed yet any molecular ions.
\end{itemize}

To summarize, the present work focused on the cold inelastic collisions inducing radiative decay between ground state Rb or Li atoms and a series of ground state ionic species (Ca$^+$, Sr$^+$, Ba$^+$, Yb$^+$). The consistent treatment of these series of pairs with the same quantum chemistry approach and the same dynamical model allows for general conclusions about the magnitude of the rate constants, and the efficiency of the formation of cold molecular ions. Dynamical results for RbSr$^+$ are published here for the first time, as well as the modeling of a heavy diatomic compound containing Yb with a large core ECP. These results could stimulate future joint experimental and theoretical investigations. For instance, the detection of a couple of shape resonances in one of these systems would put strong constraints on its electronic structure. Photodissociation experiments of the produced molecular ions such like those performed on MgH$^+$ \cite{bertelsen2004} could also provide more insight on their spectroscopy, and on their vibrational distribution (see for instance Ref.~\cite{blange1997}).

The exploration of a family of such systems like in the present work where a Rb atom interacts with an ion of a given class allows for varying -- discontinuously -- a characteristic parameter of the system, its reduced mass, revealing then the changes in the dynamics while the nature of the interactions are similar over the entire series. The same study is already achievable on the series of (Alk)-Sr$^+$ (with (Alk) = Li, Na, K, Rb, Cs) species that we previously studied \cite{aymar2011}. Based on the arguments developed in the present work, we can already draw some conclusions for this family:
\begin{itemize}
\item the LiSr$^+$ will not be favorable for the formation of deeply-bound molecular ion. Indeed, despite a noticeable TEDM of about 2a.u., the A PEC of the entrance channel -- which is almost repulsive -- has an inner turning point which coincides in position with the outer turning point of the uppermost vibrational levels of the X PEC. Moreover, the energy difference $\delta R^t_{X-A}$ is quite small, lower than 3000~cm$^{-1}$.
\item the RbSr$^+$ has been explicitly treated here, and the NaSr$^+$ and KSr$^+$ species smoothly progress toward the RbSr$^+$ with increasing  $\delta R^t_{X-A}$ concerning deeper vibrational levels
\item the CsSr$^+$ case is specific: due to the large spin-orbit in Sr($5s5p\,^3P$), the entrance channel Cs($6s$) + Sr$^+$($5s$) is located at an energy of about 20~cm$^{-1}$ and 200~cm$^{-1}$ above the Cs$^+$ + Sr($5s5p\,^3P_1$) and  the Cs$^+$ + Sr($5s5p\,^3P_0$) dissociation limits, respectively. Therefore, just like in the Rb-Ca$^+$ combination, NRCT should dominate the dynamics of this system.
\end{itemize}

Note finally that the electronic ground state of species like LiBa$^+$ and LiCa$^+$ dissociates into Li+Ba$^+$ and Li+Ca$^+$ respectively, so that in the context of hybrid trap experiments invoked here, the entrance channel is the lowest one and no loss is expected due to radiative processes. A central goal remains to progress towards lower temperature in order to approach the quantum $s$-wave collisional regime, which will be of crucial importance for the understanding and the control of such collisions which clearly departs from the semiclassical Langevin model, and for improving theoretical models.

\begin{table}[!htbp]
\caption{Energies (in cm$^{-1}$) of the quoted atom-molecule complexes in their ground state with respect to an origin of energy taken when all three atomic fragments (one (Alk) atom, one (Alk)$^+$ ion, one (Alke) atom) are at infinity. Therefore the reported quantities corresponds to the well-depths of the relevant molecular ions (from the present work) or neutrals in their electronic ground state. The Rb$_2^+$ well depth is taken from Ref.~\cite{aymar2003}, and the Li$_2^+$ (experimental) one from Ref.~\cite{bernheim1983}. }
\begin{center}
\begin{tabular}{rrrrrrr}
\hline \hline
Origin  &System     &  Energy&System     & Energy &System     & Energy  \\
\hline
Rb+Rb$^+$ + Ca&Rb + RbCa$^+$& -3851 &Ca + Rb$_2^+$& -5816 &Rb$^+$ + RbCa&-1046 \cite{pototschnig2015}\\
Rb+Rb$^+$ + Sr&Rb + RbSr$^+$& -4266 &Sr + Rb$_2^+$& -5816 &Rb$^+$ + RbSr&-1073 \cite{guerout2010,zuchowski2014}\\
Rb+Rb$^+$ + Ba&Rb + RbBa$^+$& -5292 &Ba + Rb$_2^+$& -5816 &Rb$^+$ + RbBa&-1471 \cite{gou2015} \\
Rb+Rb$^+$ + Yb&Rb + RbYb$^+$& -3435 &Yb + Rb$_2^+$& -5816 &Rb$^+$ + RbYb&-656 \cite{brue2013}\\
Li+Li$^+$ + Yb&Li + LiYb$^+$& -9383 &Yb + Li$_2^+$& -10464&Li$^+$ + LiYb&-1594 \cite{zhang2010} \\
\hline \hline
\end{tabular}
\end{center}
\label{tab:reactions}
\end{table}

\section*{Acknowledgments}
This research is supported by the Marie-Curie Initial Training Network "`COMIQ: Cold Molecular Ions at the Quantum limit"' of the European Commission under the Grant Agreement 607491. Stimulating discussions with Robin C\^ot\'e, Johannes Hecker Denschlag, Eric Hudson, and Stefan Willitsch are gratefully acknowledged. The scattering calculations have been achieved thanks to the computing facility cluster GMPCS of the LUMAT federation (FR LUMAT 2764) (http://www.gmpcs.lumat.u-psud.fr/) and to the high-performance computing center IDRIS of CNRS (http://www.idris.fr/).

\newpage

\newpage

\newpage

\newpage

\newpage

\newpage

\newpage

\newpage

\newpage



\providecommand{\newblock}{}

\end{document}